# Room temperature plasmon laser by total internal reflection

Ren-Min Ma[1*], Rupert F Oulton[1*], Volker J Sorger[1], and Xiang Zhang[1,2]

[1]NSF Nanoscale Science and Engineering Centre, 3112 Etcheverry Hall, University of California, Berkeley, CA 94720, USA.

[2]Materials Sciences Division, Lawrence Berkeley National Laboratory, 1 Cyclotron Road, Berkeley, CA 94720, USA

*These authors contributed equally to this work.

Plasmon lasers create and sustain coherent surface plasmon polaritons, collective electronic oscillations of metal-dielectric interfaces (*1-5*). These intense, coherent and confined optical fields have the unique ability to drastically enhance light-matter interactions bringing fundamentally new capabilities to bio-sensing, data storage, photolithography and optical communications (*5-7*). However, these important applications require sub-diffraction limited lasers operating at room temperature, which remains a major hurdle (*1, 2*). There are two critical challenges: high absorptive loss of metals and low cavity feedback. Recent efforts in semiconductor plasmon lasers have resulted in two approaches: devices capped in metal provide good feedback, but suffer high metal loss resulting in limited mode confinement (*2*). On the other hand, nanowire lasers on planar metal substrates achieve strong confinement with low metal loss, but the open configuration limits feedback imposing a minimum nanowire length (*1*). While the merits of capped and planar metallic devices remain mutually exclusive, plasmon lasers must rely on cryogenic temperatures to attain sufficient gain to combat losses. Therefore, room temperature plasmon laser operation below the diffraction limit demands low metal loss, effective cavity feedback and high gain; all within a single nanoscale device.

We report here a room temperature semiconductor plasmon laser with both strong cavity feedback and $\lambda/20$ optical confinement. The device consists of a 45 nm thick, 1 μm length single crystal Cadmium Sulfide (CdS) square atop a Silver surface separated by a 5 nm thick Magnesium Fluoride gap layer, shown in Fig. 1A. We achieve strong feedback by total internal reflection of surface plasmons at the cavity boundaries. The close proximity of the high permittivity CdS square and silver surface strongly confines transverse magnetic (TM) waves to the gap region with low metal loss (8), while transverse electric (TE) waves are pushed away from the metal (Fig 1A). While both mode polarizations are free to propagate in the plane, only TM modes undergo total internal reflection providing the feedback for lasing (*9,10*). This strong feedback results in sufficiently high quality factors (Q) that can be identified by cavity modes even in the spontaneous emission spectrum at low pump intensities in fig. 1B. The Q factors of the three apparent modes are 42, 82 and 69 at the resonant wavelengths of 504.8 nm, 510.3 nm and 520.2 nm, respectively, close to the predicted values in numerical simulations (*10*).

At higher pump intensities, multiple cavity modes appear with orders of magnitude higher coherence than the underlying spontaneous emission, as shown in fig. 1C. The nonlinear response of integrated output power with increasing pump intensity (inset of fig. 1C) confirms the observation of full laser oscillation. Lasing in such ultrathin devices is viable solely due the plasmonic confinement and total internal reflection feedback and was not observed in control samples consisting of CdS squares on quartz substrates where the lack of mode confinement and feedback inhibit laser action.

The intense coherent field in the gap region can interact strongly with matter (fig 1A). Such light-matter interaction enhancements are also observable in the CdS gain medium. Under weak pumping, the CdS band edge transitions of this plasmon laser show a spontaneous emission lifetime reduced by a factor of 14 (fig. 1D). The Purcell effect (*11*) is apparent in all the laser devices measured as shown in the inset of Fig 1D. Smaller devices show Purcell factors as large as 20 due to the combination of both high cavity Q and strong confinement. Larger devices have much smaller Q factors, leaving an average 2-fold reduction in lifetime due to confinement alone. Since electric field intensities in the gap region are 5 times stronger than in the CdS, we anticipate a similar increase in light-matter interaction strengths.

We have demonstrated room temperature operation of sub-diffraction limited lasers with high Purcell factors indicating their ability to strongly enhance light matter interactions. Room temperature plasmon lasers enable new possibilities in applications such as bio-sensing, data storage, photolithography and optical communications (*5-7*).

We acknowledge financial support from the US Air Force Office of Scientific Research (AFOSR) MURI program under grant no. FA9550-04-1-0434 and by the National Science Foundation Nano-scale Science and Engineering Center (NSF-NSEC) under award no. DMI-0327077.

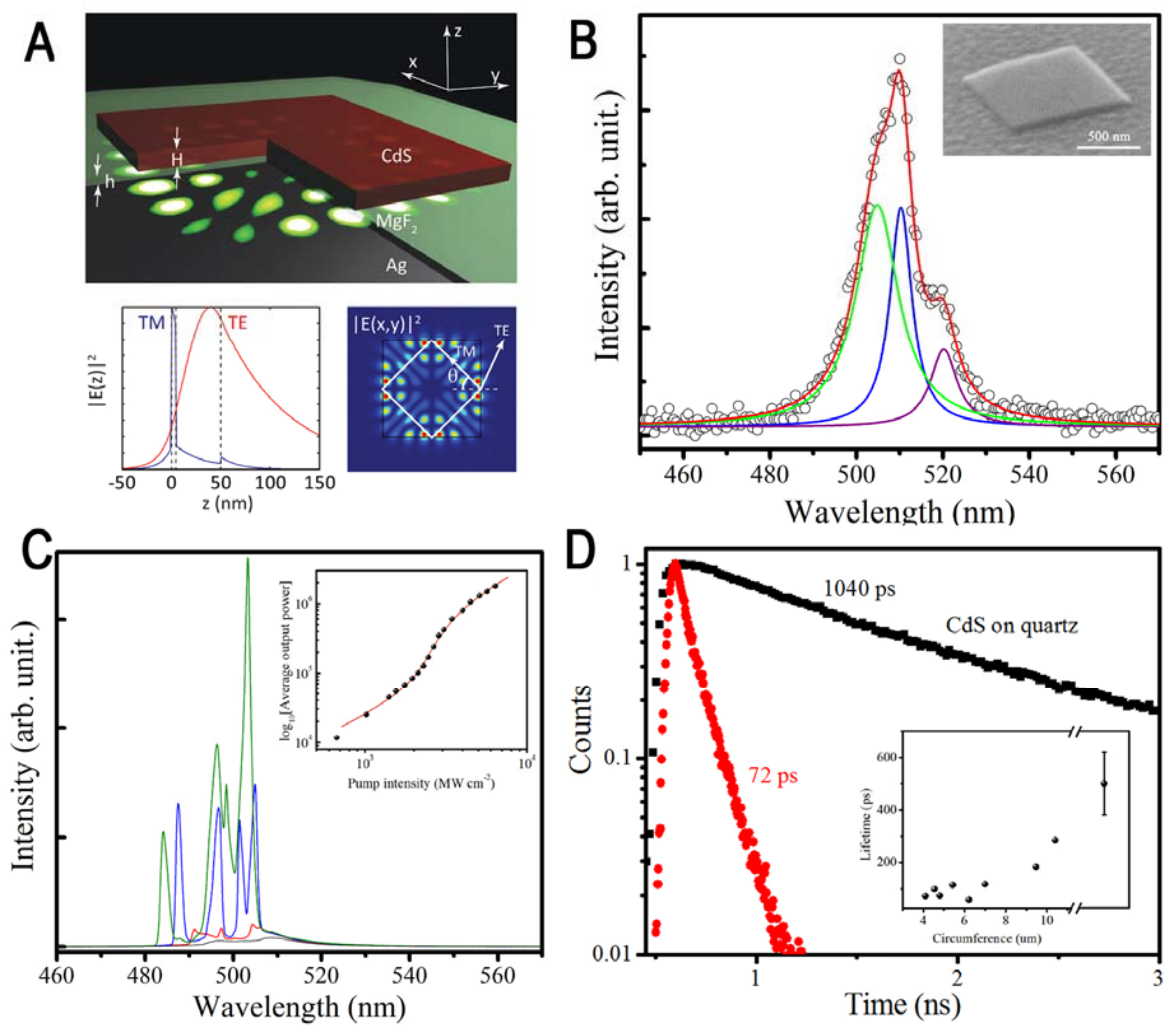

**Figure 1 A** (Top) Schematic of a room temperature plasmon laser. (Bottom-Left) The electric field distribution of TM and TE modes along with z direction. (Bottom-Right) Electromagnetic energy distribution of a TM mode in the x and y directions. **B** Room temperature (RT) spontaneous emission spectrum and de-convolved modes at low pump intensities (290 MW cm$^{-2}$) showing good cavity feedback. SEM Micrograph of the 45 nm thick, 1 μm length square Plasmon laser (inset). **C** RT laser spectra and integrated light-pump response (inset) showing the transition from spontaneous emission (1760 MW cm$^{-2}$, black) via amplified spontaneous emission (2300 MW cm$^{-2}$, red) to full laser oscillation (3070 MW cm$^{-2}$, blue; and 4020 MW cm$^{-2}$, olive) and the spectral narrowing of the laser modes. **D** Time resolved emission under weak pumping of the Plasmon laser and CdS on quartz. Inset shows decreasing lifetime of other measured devices with cavity circumference due to the Purcell effect, while very large structures exhibit a 2-fold

reduction in lifetime due to confinement along the z-direction alone. (CdS thickness varies between 45 and 88 nm.)

# Supplementary Material: "Room temperature plasmon laser by total internal reflection"

**Device fabrication and experiment set-up**

The Cadmium Sulfide (CdS) squares were made by a solution based sonication cleaving process of CdS nanobelts which were synthesized via the chemical vapor deposition method. Then the squares were deposited from solution on $MgF_2$/Ag (5 nm/800 nm) substrates. A frequency-doubled, mode-locked Ti-sapphire laser (Spectra Physics) was used to pump the squares ($\lambda_{pump}$=405 nm, repetition rate 10 KHz, pulse length 100 fs). A 20x objective lens (NA=0.4) focussed the pump beam to ~ 5 μm diameter spot on the sample. All experiments were carried out at room temperature.

**Numerical simulations**

The optical modes of the CdS square plasmonic cavities can be viewed as total internally reflected surface plasmons at the device boundaries. We verified this by comparing a 3D eigenmode analysis with a 2D+1D mode calculation. Here, we solved for the plasmonic mode in 1D and then utilized the corresponding effective mode indexes of surface waves inside and outside the cavity in a 2D eigenmode calculation. This describes the spectral positions of eigenmodes and the quality factors extremely well; i.e., the 2D simulation faithfully models the phase change of plasmonic total internal reflection and out-of-plane scattering is small.

Based on 2D+1D eigenmode simulations, we have predicted the cavity mode positions of the plasmon laser in the main text. We do this in two steps: firstly, we calculate the cavity mode dispersion for transparent CdS with various refractive indexes; we then superimpose the CdS dispersion. The cavity modes occur near the intersections as shown in Fig. S1**a**. We see good correspondence between the mode positions seen in the experiment in the region of pump powers between CdS transparency and near laser threshold (Fig. 1**b-e**).

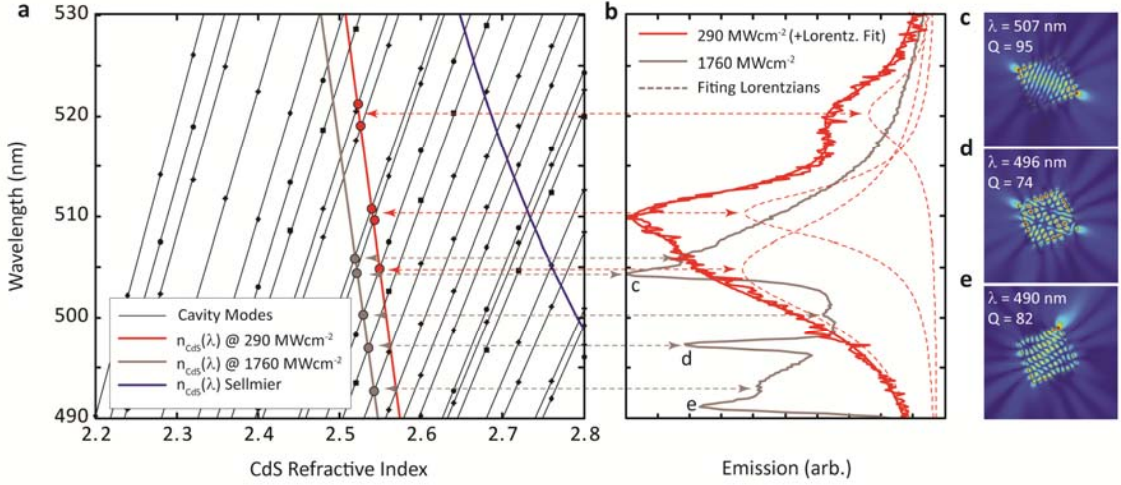

**Figure S1** Numerical simulation of the plasmon laser in the main text consisting of a 45 nm thick CdS film ($n_{CdS}$ = [2.2, 2.8]) atop a Ag film ($\varepsilon_{Ag}=\varepsilon_b-E_p^2/[E(E-i\gamma)]^{-1}$, $\varepsilon_b$ = 5, $E_p$ = 9.5 eV, $\gamma$ = 0.04 eV), separated by a 5 nm thick MgF$_2$ gap ($n_{MgF}$ = 1.38). **a** Mode dispersion as a function of CdS refractive index from 2D+1D eigenmode calculations. The intersection with the expected dispersion in CdS indicates the position of cavity modes in the experiment, shown in **b**, within the region between CdS transparency and laser threshold. **c-e** eigenmodes of the most prominent laser modes near threshold.

The discrepancy between the expected CdS dispersion from a Sellmier description ($\varepsilon_{CdS}$ = 4.23 + $\lambda^2[\lambda^2-0.18]^{-1}$) and that observed for the experiment is due to a number of factors. i) CdS dispersion changes strongly as CdS reaches transparency, ii) accuracy of gap thickness and estimated permittivity of other materials imply a level of systematic error and iii) the choice of eigenmode solver linearization point can affect the cavity mode dispersion leading to small shifts in the expected CdS dispersion.

In the case of transparent CdS, the mode quality factors are 50 < Q < 150. The majority of cavity losses arise from the metal. When metal losses are removed, Q factors increase beyond 250, indicating the excellent cavity feedback available in these devices, despite their small size.